\documentclass[aps,prd,showpacs,superscriptaddress,groupedaddress]{revtex4}%
\usepackage{graphicx}
\usepackage{dcolumn}
\usepackage{bm}
\usepackage{amssymb}
\usepackage{amsmath,amsthm}
\usepackage{amsfonts}
\usepackage{blindtext}
\usepackage{epstopdf}
\usepackage{amsmath}%
\providecommand{\U}[1]{\protect\rule{.1in}{.1in}}
\begin{document}
\title{ Newton like equations for the radiating particle: the non relativistic limit }
\author{Daniel Lluis Gonz\'alez$^{1}$, Alejandro O. Salinas$^{2}$, Alejandro Cabo
Montes de Oca}
\affiliation{\textit{Department of Theoretical Physics, Instituto de Cibern\'{e}tica,
Matem\'{a}tica y F\'{\i}sica, Calle E, No. 309, Vedado, La Habana, Cuba. } \\
\textit{$^{2}$Faculty of Physics, Havana University, San L\'azaro y L, La Habana, Cuba. }}
\date{\today}

\begin{abstract}
\noindent A broad class of forces, $P$, is identified, for which the Abraham-Lorentz-Dirac (ALD) and Newton-like equations have solutions in common. Moreover, these solutions do not present pre-acceleration or escape into infinity (runaway behavior). Any continuous or piecewise continuous force can be represented in terms of functions belonging to this class $P$. It was also argued that the set of common solutions of both sets of equations is wider, it was proved that these solutions could be formulated in terms of generalized functions. The existence of such generalized functions motions is explicitly demonstrated  for the relevant example of the instantaneously applied constant force,  for which the respective solution of the ALD equation  exhibits lack of causality and runaway motion. In this case, the expressions for the position and velocity of the particle are formulated in terms of generalized functions of time,  having only a  point support at the time that the force is applied. Thus, both, the velocity and the position are discontinuous at the instant that the force was applied. The solution for a  particle moving between the plates of a capacitor  reproduces the one obtained by A. Yaghjian from his discussion for extended particles. This outcome suggests a possible link or equivalence between both studies. A solution, common to the Newton-like and the ALD equations, for a constant homogeneous magnetic field is also presented. The extension of the results obtained to a relativistic context will be studied in future works.

\end{abstract}

\pacs{03.50.De, 02.30.Ks, 41.60.-m, 24.10.Cn}
\maketitle



\section{ Introduction}

It has been known for more than 100 years that an accelerated charge radiates,
and that the effects of this radiation on the charge itself can be described as a reaction force that acts as a friction on the particle. Surprisingly, also during this last century, a great effort has been put to find an appropriate equation to describe the motion of a charged particle. Great physicists,
starting with Abraham and Lorentz, have worked on this issue and already today
there are hundreds of articles in the literature dealing with the topic
\cite{Lorentz} - \cite{PRD}. The ALD equations were introduced more than a
century ago to describe the interaction of a charged particle with its own electromagnetic field and its relativistic version was developed by Dirac in 1938 \cite{Dirac}.

The treatment of this problem by classical electromagnetism
has been plagued with difficulties, although this theory has been used to obtain
equations that very accurately describe radiation for the most
important applications. Starting from Maxwell's equations for the particle's field and Newton's equations for the particle an equation of motion was obtained, one that included the forces produced by the particle's radiated field. This system of equations proved to be suitable for the description of a macroscopic particle, yet not for a point particle or even for small enough particles. In the low speed limit, these equations are called the non-relativistic ALD equations. These equations incorporate the effects of the radiation produced by this classical charged particle on the particle itself, that is, they include a term for the reaction force due to the radiation. The fact that these equations of motion are not of second order, that is, there is a third order derivative of position involved, leads to fundamental issues for the solution, such as non-causality, runaway behavior or contradictions with Special Relativity. One of the main reasons behind such problems is that the punctual charge's field, supposedly Coulombian, diverges around the position of the particle itself. This situation causes the need for a renormalization
of the mass and the existence of the \emph{runaway} solutions. Nevertheless, the works of Poincar\'e and Dirac solved the contradiction with special relativity but the
other problems remained \cite{Dirac}.

Some authors maintain that these difficulties  can be avoided if the assumption of
the electron as a point particle is eliminated, because if the gravitational
effects are negligible, then, a point particle with finite charge and mass is
impossible in classical physics. For example, in references \cite{Ford, oconnell} the authors affirm that the particle should be considered as extended and they derive a second-order equation of motion, which they claim to be exact, assuming the existence of an internal structure within the charged particle. Reference \cite{Yighjian} is another work that sustains this line of thought, arguing that for forces that do not depend analytically on time, the ALD equations cannot be derived exactly from the
coupled equations for the particle reacting to its own electromagnetic field. Likewise, in reference \cite{Raju} the authors
introduce a special form of the extended structure that also removes the
\emph{runaway} behavior and pre-acceleration.

Also, in a different line of thought, some authors maintain that certain physically-coherent initial conditions should be imposed upon the solutions of the equation of motion. Some of them state that this procedure could provide an exact
equation. This idea was originally proposed by Dirac himself \cite{Dirac}. In the
article \cite{Spohn}, and some of its references, this point of view has been thoroughly examined. In these works, special constraints have been proposed in order to prevent undesirable properties to show up in the solutions of the ALD equations.

Another variation of the ALD equations consists in developing a solution through several iterations, and generalizing each term into a series. This equation is known as the Landau-Lifshitz one, which has been carefully
studied by some authors.

However, an also numerous group of authors consider  the ALD equations as rigorously valid. Therefore, it can be concluded that there is still a heated discussion about the ALD equation.

In reference \cite{PRD}, a discussion  was presented that can be considered
within the just mentioned line of thought. A Newton-like set of equations  was examined, whose solutions also solved the ALD equations. In addition, a covariant generalization
of these Newton-like equations was proposed, which motions also satisfy  the ALD equations. Although the proposed Newton-like equations suggest that there is no pre-acceleration or runaway in their solutions, the relationship between the two kinds of solutions, those of the Newton-like equations and those of the ALD ones, wasn't clear from the beginning. In particular, it was not understood if the ALD equations have or not more physically coherent solutions than the Newton like ones. This question is still open and is one of the main issues that will be addressed in further works. Its understanding will define if the solutions of the Newton like equations  contain or not the whole spectrum of possible motions of the  radiating particle.

The present work begins the study of the connections between the solutions of
the two types of equations. We start here with their non-relativistic version,
which deals with fewer conceptual difficulties. The results of this first study
are expected to be helpful in understanding the relationship between the
relativistic forms of both equations.  In addition, we will concentrate in generalizing the discussion to include solutions in terms of generalized functions. Then, we will study the motion for an instantaneously applied force described by Heaviside's unitary step function, which is the most relevant case that shows non-causal and runaway behavior in this context.

Our first goal is to introduce a class of forces for which the Newton-like equations are convergent, and therefore, well defined. This set consists of all forces $f^{i}(t)$ represented by polynomials of the time $t$. These functions then form the space $P$.
Therefore, by virtue of the Weierstrass approximation theorem, it follows that
with functions of this kind, it becomes possible to approximate any continuous
or piecewise continuous force as much as it is desired in any closed time interval.

Next, another goal of this study is to formulate non-relativistic solutions common to both equations in terms of generalized functions. For that, we will employ functions of the space $P$.
Then, we will explicitly verified that these solutions are causal and do not show runaway behavior. The solvability of these equations is checked in the sense of generalized functions, according to the theoretical framework introduced by Colombeau and Egorov in \cite{colombeau} - \cite{egorov}. This modern approach to generalized functions can be applied even to non-linear problems, in contrast to the more classical theory of Schwartz and Sobolev \cite{colombeau}.

Further, we focus on to explicitly solve the particular case of the
instantaneously applied constant force. An exact formula is obtained for the
coordinates and velocities, once the limits of the sequences that define the solutions are calculated for all times except the instant the force is applied. It follows, that the coordinates and the velocities of the particle  have discontinuous jumps at the instant in
which the force is applied. The absence in the literature of this kind of
solution is justified, since it violates the usual rule that the coordinates
of the particle do not suddenly change. In fact, this is not the case, since forces depending on derivatives of the Dirac Delta function show up. These kinds of forces clearly lead to  discontinuous jumps in the initial
coordinates of the particle. Moreover, this property represents
a contradiction with special relativity, since it requires speeds greater than the speed of light. However, in the framework of the non-relativistic movement
under consideration, this property still doesn't represent a conflict. Then, it is in our best interest to verify that there exists a relativistic solution whose limit, when the speed of light becomes infinite, leads to the non-relativistic solution. This is expected to be studied in subsequent works.

It can be helpful here to further insist in the  physical relevance of
the forces represented by Heaviside's function.   For example, they are  relevant in particle
accelerators, in all kinds of experiments in  which many charged particles suddenly enter
regions where electric or magnetic forces exist. Thus, the effective forces
acting on them can be accurately represented as instantaneously applied and constant for short time periods. For these reasons this simple example has been frequently discussed when the ALD equations are studied. As a matter of fact, it is precisely for these forces
that some solutions of the ALD equations predict non physical results for
the motion: runaway solutions and lack of causality. It is mainly because of this, that this example will have a central role in this work.

However, as it was stated before, we intend to present a general discussion about how to define a solution in terms of generalized functions, specifically to address the problem of the force represented by a Heaviside function. Moreover, the solution we developed was causal and did not show runaway behavior.

Actually, the aforementioned problems were not discussed in reference \cite{PRD}, as no solutions were formulated in the sense of generalized functions. Although, that initial analysis was required to proceed to the study of the problem with a force represented by a Heaviside function.

We estimate that this study explores from a new perspective, the difficult  problem of
finding consistent solutions to the ALD equations, that do not show any kind of  non-causal or runaway behavior. These solutions should be easier to find in a non-relativistic context, as a first approach to this problem. This problem has already been solved, at least for the non-relativistic ALD and the Newton-like equations. Moreover, since we now included solutions formulated in terms of generalized functions, there are now more solutions available for both equations and so, it seems more likely to find properly consistent solutions.

It is of utter importance in this work that the solutions found for a particle moving between the two plates of an electric capacitor that produces a constant electric field, coincide with the ones in \cite{Yighjian}.
The developed non-relativistic solution allows us to directly evaluate the energy and momentum radiated by the particle at the input and the output of the capacitor. This result suggests that the analysis presented here
is perhaps equivalent to the general discussion developed in reference
\cite{Yighjian}. Also, in that work, it was proven that if the spatial extension of the particle is taken into account, the ALD equation ceases to be valid. We expect to assess the possible equivalence between the analysis done here and that of reference \cite{Yighjian} once the results of this work are extended to the relativistic context.

We also present the exact solution of the Newton-like equations for the case
of an homogeneous magnetic field. This solution also satisfies the non-relativistic ALD equation.

The work is structured as follows. In Section II, it is checked that well defined solutions of the Newton-like equations are also exact solutions of the non-relativistic ALD equation. Thereafter, in section III, a new class of solutions for the Newton-like equation is identified, that is, a kind of solutions that can be represented by polynomial functions of time. Then, it is in section IV that the main result of this work is presented: there are solutions for both ALD and the Newton-like equations that can be expressed in terms of successions of functions defined by these aforementioned polynomials of the time $t$, in the modern sense of generalized functions according to Colombeau and Egorov \cite{colombeau} - \cite{egorov}. In section V, the existence of this kind of solutions is proven with an example relevant to the subject, that of the suddenly applied constant force. Further, in section VI, the solution for a charge moving between the plates of an electric capacitor is developed. Moreover, this solution turned out to be the very same that A. Yaghjian developed in  \cite{Yighjian} for a radiating extended charge. Finally, in section VII, the solution of the Newton-like equation for an homogeneous magnetic field is found. Then, it is shown that this solution satisfies the ALD equations as well.

\section{ The Newton-like equations}

Let us consider a non-relativistic motion of a particle through space-time
described by a trajectory $C$ defined by the curve $x(t)=\left(
x^{1}(t),x^{2}(t),x^{3}(t)\right)  $ and parameterized by time $t$. All along this work we will employ natural units, in which distance and time are measured in $cm$ and the mass in $cm^{-1}$.

In this case, the ALD equations are:
\begin{align}
\label{ald}ma^{i}(t)-f^{i}(t)  &  =k\frac{da^{i}(t)}{dt},\\
v^{i}(t)=\dot{x^{i}}(t)  &  =\frac{dx^{i}(t)}{dt},\\
a^{i}(t)=\dot{v^{i}}(t)  &  =\frac{dv^{i}(t)}{dt},
\end{align}
where the index $i$ has the three values $i=1, 2, 3$.

Then, a class of exact solutions of equation (\ref{ald}) is defined by
the so-called Landau-Lifshitz series
\cite{jackson, PRD}. Below, we will refer to these equations as the
Newton-like equations, explicitly defined as%
\begin{align}
\label{aceleracion}a^{i}(t)=\frac{1}{m} \sum_{n=0}^{\infty} \frac{d^{n}%
}{dt^{n}}f^{i}(t)\left(  \frac{k}{m}\right)  ^{n}.
\end{align}
That the solutions of the equations (\ref{aceleracion}) also solve the ALD
ones can be easily checked considering that%
\begin{align}
\dot{a^{i}}(t)  &  =\frac{1}{m} \sum_{n=0}^{\infty} \frac{d^{n+1}}{dt^{n+1}%
}f^{i}(t)\left(  \frac{k}{m}\right)  ^{n}\nonumber\\
&  =\frac{1}{k} \sum_{n=0}^{\infty} \frac{d^{n+1}}{dt^{n+1}}f^{i}(t)\left(
\frac{k}{m}\right)  ^{n+1}\nonumber\\
&  =\frac{1}{k} \left[  \sum_{n=0}^{\infty} \frac{d^{n}}{dt^{n}}%
f^{i}(t)\left(  \frac{k}{m}\right)  ^{n}-f^{i}(t)\right] \nonumber\\
&  =\frac{1}{k} \left[  ma^{i}(t)-f^{i}(t)\right]  .
\end{align}

This form of the Landau-Lifshitz series is a valid solution for accelerations small enough to be considered non-relativistic, as it is stated in references \cite{jackson} and \cite{zhang}.
However, the Landau-Lifshitz series may  diverge and constitutes an asymptotic series of the solution of the ALD equation. It should be remarked, that, up to our knowledge, a proof does  not exist requiring  that the Landau-Lifshitz series should converge for a  force to be physically meaningful.
If such a condition results  in fact valid, it constitutes a proof  of the  equivalence between the  ALD and Newton like equations. The
important point for us in what follows is that, assuming that the series is
well defined, the path $x^{i}(t)$ exactly satisfies the non-relativistic ALD equation:%

\begin{align}
\label{ald1}ma^{i}(t)-f^{i}(t)=k \dot{a}(t).
\end{align}

\section{A wide class of solutions for the Newton-like equations}

The aforementioned Newton-like equation for the non-relativistic ALD equation
has the form%

\[
a^{i}(t)=\frac{1}{m}\sum_{n=0}^{\infty}\frac{d^{n}}{dt^{n}}f^{i}(t)\left(
\frac{k}{m}\right)  ^{n}.
\]

Note that the name: Newton-like, means that the equation has the form of a Newtonian equation in which the acceleration is defined by a series.
However, the operator $\widehat{O}$%

\begin{align}
\widehat{O}f^{i}(t)=\frac{1}{m}\sum_{n=0}^{\infty}\frac{d^{n}}{dt^{n}}%
f^{i}(t)\left(  \frac{k}{m}\right)  ^{n},
\end{align}
may not converge for a wide class of functions. Therefore, the answer to the
following question is not clear: Can any force $f^{i}(t)$ , for example: a
continuous one, be linearly approximated by functions belonging to the class
for which the series converges?

The presence of arbitrary time derivatives in the Newton-like equation
directly indicates that if the components of the force $f^{i}(t)$ are
polynomial of the time variable $t$, the series will be always convergent and
the finite expression obtained will be an exact solution of the ALD equation.
We will call the set of all polynomial functions of time defined in an
arbitrary and closed interval $(a, b)$ as $P$.

The number of possible solutions in this class of functions is large. This is a direct consequence of the so-called  Weierstrass Approximation Theorem: \emph{Any
continuous function on the interval $(a, b)$ can be uniformly approximated by
polynomials on this interval}.

This theorem implies that it is possible to arbitrarily approximate  continuous
forces by functions belonging to $P$. In addition to this, the theorem has been
generalized for piecewise continuous functions like the Heaviside's unitary step function.

\section{ Formal solution in terms of generalized functions}

\ As the class of functions P seems to be general enough, the formulation of solutions common to the Newton-like and the ALD equations within the set of continuous (or piecewise continuous) forces seems feasible. The solutions
appear as successions of functions, that satisfy the ALD equations in the
sense of the generalized functions and belong to the set $P$. That is,  the ALD equations after multiplied by a test function (infinitely differentiable
and with compact support) and integrated, vanish, after taking the limit of
the index of their respective associated succession to infinity.

Let us first define a succession of functions, belonging to the previously defined set $P$, as follows. We will assume that the force $f_{i}(t)$ is a
continuous (or piecewise continuous) function of time and that it can have a
divergent Landau-Lifshitz (LL) series for certain values of the time. Then,
$f^{i}(t)$ can be approximated around the instant in which the LL
series diverges, with arbitrary precision due to the Weierstrass
approximation theorem as%

\begin{equation}
\lim_{n\rightarrow\infty}f_{n}^{i}(t)=f^{i}(t),
\end{equation}
where the expansion $f_{n}^{i}(t)$ is a polynomial function in $t$ pertaining to
$P$. Let us now consider a force with a LL series showing divergent behavior at time $t$ and, as well, let us assume thet the interval $a\leq t\leq b$ is completely included
in the larger interval $0<t<1.$

Let us further impose that $\alpha$ and $\beta$ are such that%

\begin{equation}
0<\alpha<a<b<\beta<1.
\end{equation}

Then, the arbitrarily precise polynomial approximation of the force $f^{i}(t)$
can be expressed in the form defined in reference \cite{courant} by the
formula%
\begin{equation}
f_{n}^{i}(t)=\frac{\int_{\alpha}^{\beta}f^{i}(s)\left[1-(s-t)^{2}%
\right]^{n}ds}{\int_{-1}^{1}\left[1-(s-t)^{2}\right]^{n}ds}. \label{fi}%
\end{equation}
\newline

This representation can be conceived as the integral of the force
multiplied by a very localized  contributions to the force over all the space in which the force is defined. Intuitively, this can actually be interpreted as an integral over all the punctual components of the force over the range of definition, ergo, as the sum of infinitely many contributions of Dirac Delta functions of the argument $(t-s)$. \newline

\begin{figure}[h]
\begin{center}
\includegraphics[width=0.5\textwidth]{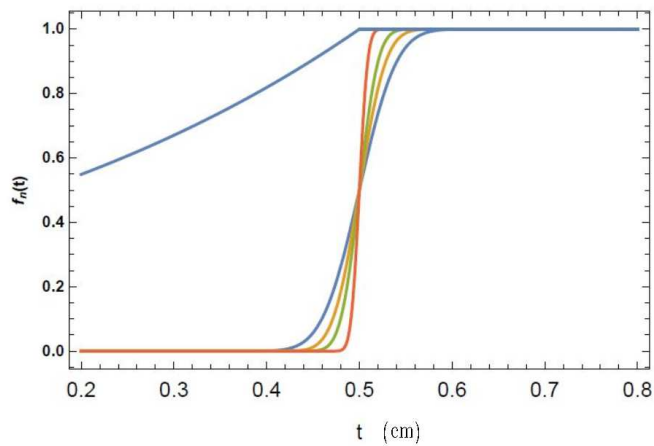}
\end{center}
\caption{The figure shows four elements of the sequence that approximates the
constant force applied at time $t=\frac{1}{2}$ $cm$ (the time and distance are measured in natural units cm), for the index values: $n=500,
1000, 2000, 10000$. The effective force associated with the non causal but
controlled solution of the ALD equation is also plotted in blue.}%
\label{fn}%
\end{figure}

\subsection{ The existence of generalized solutions}

Let us now consider the aforementioned formal proof that solutions in the sense of
the generalized functions of the ALD equations can exist as sequences of
functions defined in the space $P$.

The satisfaction of the ALD equation in the sense of the Colombeau-Egorov
theory of generalized functions \cite{colombeau}-\cite{egorov} in the internal
region of the interval $(a,b)$ can be shown if it is possible \ to define in
$P$ a succession of coordinates, $x_{n}^{i}(t)$, $n=1,2,...\infty$ for which
the following limit vanish
\begin{align}
E  &  =lim_{n\rightarrow\infty}\int_{a}^{b}\phi(t)\left[  m(1-\tau\partial
_{t})a_{n}^{i}(t)-f_{n}^{i}(t)\right]  dt\nonumber\\
&  =0,
\end{align}
for arbitrary values of the test functions of time $\phi(t)$ (infinitely
differentiable and with compact support contained in the interval $(a,b))$.
The other quantities related to the equations above are the velocities and
accelerations
\begin{align*}
v_{n}^{i}(t)  &  =\frac{d}{dt}x_{n}^{i}(t),\\
a_{n}^{i}(t)  &  =\frac{d}{dt}v_{n}^{i}(t).
\end{align*}

Then, we can define the following sequence of functions:

\begin{equation}
a_{n}^{i}(t)=\frac{1}{m}\sum_{l=0}^{N_{n}}\tau^{i}\frac{d^{l}}{dt^{l}}%
f_{n}^{i}(t),\text{ \ \ }n=1,2,3,... \label{aceleracionk}%
\end{equation}
formulated in terms of the force functions $f^{i}_{n}(t)$. Also, $N_n$ is an integer that tends to infinity when $n$ also does and it fulfills the condition:
\begin{equation}
N_{n}+1>2n. \label{Nn}%
\end{equation}
Below, we will argue  that the succession  (\ref{aceleracionk}) satisfies the ALD equations in the sense of the generalized functions
\cite{colombeau}-\cite{egorov} in the limit $n\rightarrow\infty$.

Substituting (\ref{aceleracionk}) in the solvability condition, we obtain

\begin{align}
E  &  =lim_{n\rightarrow\infty}\int\phi(t)\left[  (1-\tau\partial_{t}%
)\sum_{m=0}^{N_{n}}\tau^{m}\frac{d^{m}}{dt^{m}}f_{n}^{i}(t)-f_{n}%
^{i}(t)\right]  dt\nonumber\\
&  =lim_{n\rightarrow\infty}\int-\phi(t)\tau^{Nn+1}\frac{d^{N_{n}+1}%
}{dt^{N_{n}+1}}f_{n}^{i}(t)dt\nonumber\\
&  =0, \label{heavisidex}%
\end{align}
which vanishes due to relation (\ref{Nn}),  since the only derivative appearing acting on the polynomials $f_{n}^{i}(t)$ is zero. That is, the number of
derivatives is greater than the order of the polynomial on which the
derivatives act in the integrand of (\ref{heavisidex}). Therefore, the defined
sequence satisfies the ALD equations in the sense of generalized functions.

Let us now show that the Newton-like equations satisfy their corresponding solvability condition in the sense of generalized functions. After substituting (\ref{aceleracionk}) into that equivalent condition, we get:

\begin{align}
E  &  =lim_{n\rightarrow\infty}\int\phi(t)\left[  \sum_{m=0}^{N_{n}}\tau
^{m}\frac{d^{m}}{dt^{m}} f_{n}^i(t) -a_{n}^{i}(t) \right]  dt\nonumber\\
&  =0. \label{heaviside}%
\end{align}
which vanishes directly due to the same definition of the  sequences (\ref{aceleracionk}).
 In this case the integrand in
(\ref{heaviside}) is exactly zero if the sequences  turn to be properly
defined for all the sequence indices and parameters.

Thus, the existence of solutions to the ALD and the Newton-like equations in the form of infinite successions of functions defined in the space $P$ seems plausible. Although, for them to effectively exist, it is required
that the  infinite sequences should be also well defined for all the
sequence indices and parameters before taking the limit. This   condition  seems to be assured by the Weierstrass theorem.

Finally, we will prove in the next section that such generalized functions actually exist. To support this claim, we will study the particular case of a suddenly applied constant force. It can be underlined
that this systems had been a relevant primer in connection with the
discussions of the properties of radiating particles along the times.

\section{ The instantaneously applied constant force}

Let us now consider the example of the instantaneously applied constant
force. As commented before, the resolution of this problem will conclude our analysis in the present paper on the generalized functions of the space $P$.

Let's assume that the force is applied at a time $t=\frac{1}{2}$. Then, this force can be represented by a Heaviside unitary step function. Now, let us shift the argument of that unitary step function by $\tau=\frac{1}{2}$, then:
\begin{align*}
f^{i}(t)  &  =f^{i}\overline{\theta}\left(  t-\frac{1}{2}\right) \\
&  =f^{i}\theta(t),
\end{align*}
where  $\overline{\theta}(t)$ is the usual Heaviside function.%

\begin{equation}
\overline{\theta}(t)=\left\{
\begin{tabular}
[c]{l}%
$0$ \ \ \ for $t<0$\\
$1$ \ \ \ for $\ t\geq0$%
\end{tabular}
\ .\right.
\end{equation}

The constants $a, b, \alpha, \beta$ will be chosen with the values%
\begin{align}
a  &  =0.001,\\
\alpha &  =0.002,\\
\beta &  =0.998,\\
b  &  =0.999.
\end{align}

Then, integrating the expression of the acceleration succession, the formula for the velocity associated with the solution $a_{n}^{i}(t)$, in the form of a succession in the interval $(a, b)$ (after taking as initial
condition the velocity $v_{0}^{i}$ at an instant $t_{0}$ $<\frac{1}{2}$) becomes%

\begin{align}
v_{n}^{i}(t)  &  =v_{0}^{i}+\int_{t_{0}}^{t}a_{n}^{i}(t^{\prime})dt^{\prime
}\nonumber\\
&  =v_{0}^{i}+\frac{1}{m}\sum_{l=0}^{N_{n}}\tau^{l}\int_{t_{0}}^{t}\frac
{d^{l}}{d{s}^{l}}f_{n}^{i}(s)ds\nonumber\\
&  =v_{0}^{i}+\frac{f^{i}}{m}\int_{t_{0}}^{t}\theta_{n}(t^{\prime})dt^{\prime
}+\frac{f^{i}}{m}\tau\left[  \theta_{n}(t)-\theta_{n}(t_{0})\right]
+\label{v}\\
&  \frac{f^{i}}{m}\tau^{2}\delta_{n}\left(  t-\frac{1}{2}\right)  +\frac
{f^{i}}{m}\sum_{l=3}^{N_{n}}\tau^{l}\left[  \frac{d^{l-2}}{d{t^{\prime}}%
^{l-2}}\delta_{n}\left(  t^{\prime}-\frac{1}{2}\right)  \right]
\bigg|_{t^{\prime}=t}.\nonumber
\end{align}
where we defined%

\begin{align}
\frac{d}{dt}\theta_{n}(t)  &  =\frac{d}{dt}\overline{\theta}_{n}\left(
t-\frac{1}{2}\right) \nonumber\\
&  =\delta_{n}\left(  t-\frac{1}{2}\right)  ,
\end{align}
in which $\delta_{n}(t-\frac{1}{2})$ is a polynomial regularization of the
Dirac Delta centered at $t=\frac{1}{2}.$\newline

Integrating again and assuming the initial condition for the coordinate as
$x_{0}^{i}$ we have for the coordinates the expression%

\begin{align}
x_{n}^{i}(t)  &  =x_{0}^{i}+v_{0}^{i}(t-t_{0})+\int_{t_{0}}^{t}v_{n}%
^{i}(t^{\prime})dt^{\prime}\nonumber\\
&  =x_{0}^{i}+v_{0}^{i}(t-t_{0})+\frac{f^{i}}{m}\sum_{l=0}^{N_{n}}\tau^{l}%
\int_{t_{0}}^{t}d{t^{\prime\prime}}\int_{t_{0}}^{{t^{\prime\prime}}}%
\frac{d^{l}}{d{t^{\prime}}^{l}}\theta_{n}(t^{\prime})dt^{\prime}\nonumber\\
&  =x_{0}^{i}+v_{0}^{i}(t-t_{0})\label{x}\nonumber\\
&  +\frac{f^{i}}{m}\int_{t_{0}}^{t}d{t^{\prime\prime}}\int_{t_{0}}%
^{{t^{\prime\prime}}}\theta_{n}(t^{\prime})dt^{\prime}+\frac{f^{i}}{m}\tau
\int_{t_{0}}^{t}[\theta_{n}(t^{\prime})-\theta_{n}(t_{0})]dt^{\prime}+\nonumber\\
&  \frac{f^{i}}{m}\tau^{2}[\theta_{n}(t)-\theta_{n}(t_{0})]+\frac{f^{i}}%
{m}\sum_{l=3}^{N_{n}}\tau^{l}\left[  \frac{d^{l-3}}{d{t^{\prime}}^{{l-}3}%
}\delta_{n}\left(  t^{\prime}-\frac{1}{2}\right)  \right]
\bigg|_{t^{\prime}=t}.
\end{align}

The terms of higher order in the derivatives in the respective equations for
the velocity and the coordinates have the forms

\begin{eqnarray}
T_{v}^{n}(t)  &  = &\frac{f^{i}}{m}\sum_{l=3}^{N_{n}}\tau^{l}\left[
\frac{d^{l-2}}{d{t^{\prime}}^{l-2}}\delta_{n}\left(t^{\prime}-\frac{1}{2}\right)\right]\bigg|_{t^{\prime}=t},\\
T_{x}^{n}(t)  &  = &\frac{f^{i}}{m}\sum_{l=3}^{N_{n}}\tau^{l}\left[
\frac{d^{l-3}}{d{t^{\prime}}^{l-3}}\delta_{n}\left(t^{\prime}-\frac{1}{2}\right)\right]\bigg|_{t^{\prime}=t},
\end{eqnarray}
where $N_{n}$ can be substituted by infinity, because for every $l>N_{n}$ the
number of derivatives present in both relations acting on the polynomials,
that define the functions $\delta_{n}(t^{\prime})$, make all these terms vanish. But, taking then the limit $n\rightarrow\infty$, for all times $t$ different than the moment in which the force was applied $t=\frac{1}{2}$, we have:

\begin{align}
\lim_{n\rightarrow\infty}T_{v}^{n}(t)  &  =\frac{f^{i}}{m}\sum_{l=3}^{\infty
}\tau^{l} \left[  \frac{d^{l-2}}{d{t}^{l-2}}\lim_{n\rightarrow\infty}\delta
_{n} \left(  t-\frac{1}{2} \right)  \right]\bigg|_{t\neq\frac{1}{2}%
}\nonumber\\
&  =\frac{f^{i}}{m}\sum_{l=3}^{\infty}\tau^{l} \left[  \frac{d^{l-2}}%
{d{t}^{l-2}}\delta \left(  t-\frac{1}{2} \right)  \right]\bigg|_{t\neq\frac{1}{2}}\nonumber\\
&  =0,\\
\lim_{n\rightarrow\infty}T_{x}^{n}(t)  &  =\frac{f^{i}}{m}\sum_{l=3}^{\infty
}\tau^{l} \left[  \frac{d^{l-3}}{d{t}^{l-3}}\lim_{n\rightarrow\infty}\delta
_{n} \left(  t-\frac{1}{2} \right)  \right]\bigg|_{t\neq\frac{1}{2}}%
,\nonumber\\
&  =\frac{f^{i}}{m}\sum_{l=3}^{\infty}\tau^{l} \left[  \frac{d^{l-3}}%
{d{t}^{l-3}}\delta \left(  t-\frac{1}{2} \right)  \right]\bigg|_{t\neq\frac{1}{2}}%
,\nonumber\\
&  =0.
\end{align}

That is, except for the time in which the force was applied $t=\frac{1}{2}$, the corrections to the velocities and coordinates determined by these two terms
are zero, because all the derivatives of the Delta function vanishes outside their
point support. Therefore, the following expressions for the velocity and the
coordinates fully determine the values of these two magnitudes at all times
except for the moment in which the force is applied $t=\frac{1}{2}$%

\begin{eqnarray}
v^{i}(t) & = & v_{0}^{i}+\frac{f^{i}}{m}\int_{t_{0}}^{t}\theta(t^{\prime
})dt^{\prime}+\frac{f^{i}}{m}\tau\text{ }\theta(t)+\frac{f^{i}}{m}\tau
^{2}\delta\left(  t-\frac{1}{2}\right)  ,\label{vf}\\
& = & v_{0}^{i}+\frac{f^{i}}{m}\text{ }t\text{ }\theta(t)+\frac{f^{i}}{4\text{
}m}\delta\left(  t-\frac{1}{2}\right) , \\
x^{i}(t) & = & x_{0}^{i}+v_{0}^{i}(t-t_{0})+\frac{f^{i}}{m}\int_{t_{0}}%
^{t}d{t^{\prime\prime}}\int_{t_{0}}^{{t^{\prime\prime}}}\theta(t^{\prime
})dt^{\prime}+\frac{f^{i}}{m}\tau\int_{t_{0}}^{t}[\theta(t^{\prime}%
)-\theta(t_{0})]dt^{\prime}+ \noindent\\
&  &\frac{f^{i}}{m}\tau^{2}[\theta(t)-\theta(t_{0})]\label{xf} \noindent \\
& = & x_{0}^{i}+v_{0}^{i}(t-t_{0})+\frac{f^{i}}{2m}\left(t-\frac{1}{2}\right)^{2}%
\theta(t)+\frac{f^{i}}{2m}\left(t-\frac{1}{2}\right)\theta(t)+\frac{f^{i}%
}{4m}\theta(t),\\
& = & x_{0}^{i}+v_{0}^{i}(t-t_{0})+\frac{f^{i}}{2m}\theta(t)\left(  t^{2}+\frac{1}{4}\right),
\end{eqnarray}
in which it was considered that $\theta(t_{0})=0$.

Hence, the solution defined by the velocity $v^{i}(t)$ and coordinates $x^{i}(t)$, which are taken in the limit $n\rightarrow\infty$ at all times $t\neq\frac{1}{2}$, is well defined at every instant except for the moment the force was applied.

Let us illustrate the above result with an example, in order to clarify how the solutions in terms of generalized functions can be approximated by successions at every instant of time, except for the moment the force was applied. As it is difficult to plot the sum of all the terms, let's take the following approximation for the acceleration%

\begin{equation}
a_{n}^{(3)i}(t)=\frac{1}{m}\sum_{l=0}^{3}\tau^{i}\frac{d^{l}}{dt^{l}}f_{n}%
^{i}(t). \label{a3}%
\end{equation}

and to simplify the numbers, the ratio between the force and the mass will take
the value

\begin{equation}
\frac{f}{m}= 1,
\end{equation}

Also, we will assume that the motion is in one dimension, therefore, we will only consider the component $i=1$.

\begin{figure}[h]
\begin{center}
\includegraphics[width=0.5\textwidth]{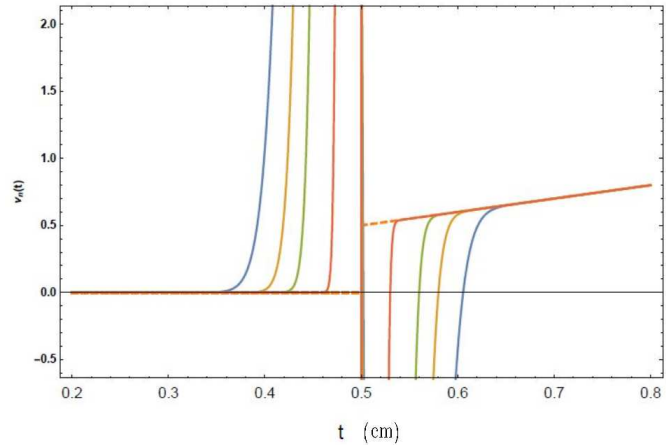}
\end{center}
\caption{ The iterative solutions for the velocity $v_{n}(t)$ in the approximation
of the acceleration taken $a^{(3)}_{n}(t)$ in (\ref{a3}) as functions of the time  in $cm$ (the time and distance are measured in natural units cm). Notice how, as the
index $n$ of the succession associated to the solution in the form of a
generalized function grows, the coordinates get closer and closer to the
expression (\ref{vf}). The curves are reducing their width corresponding to
the growth of their index $n$ in the values $n=500, 1000, 2000, 10000$.}%
\label{vn}%
\end{figure}
That is, to illustrate this solution we will omit the terms of order higher than 3 in
the powers of $\tau$ and in the derivatives. More terms could be included, but
what they produce is a greater complexity in the curves in the area very close
in time to the point of connection of the force.
Nevertheless, these terms vanish in the limit $n\rightarrow\infty$ for every instant $t\neq\frac{1}{2}$.

The figure \ref{vn} shows the expressions for the velocity that is obtained by
integrating over time the terms of the successions of accelerations
$a_{n}^{(3)i}(t)$ for a one-dimensional problem ($i$ takes a single value
$i=1).$ The values of the indices of the sequence and the set of parameters are%

\begin{align}
n  &  =500,1000,2000,10000,\\
m  &  =1,\text{ \ }f^{i}=1,\text{ \ }i=1.
\end{align}

For the initial conditions for the velocity and the coordinate the following values were chosen

\begin{align}
v(-1)  &  =0,\\
x(-1)  &  =0.
\end{align}
\begin{figure}[h]
\begin{center}
\includegraphics[width=0.5\textwidth]{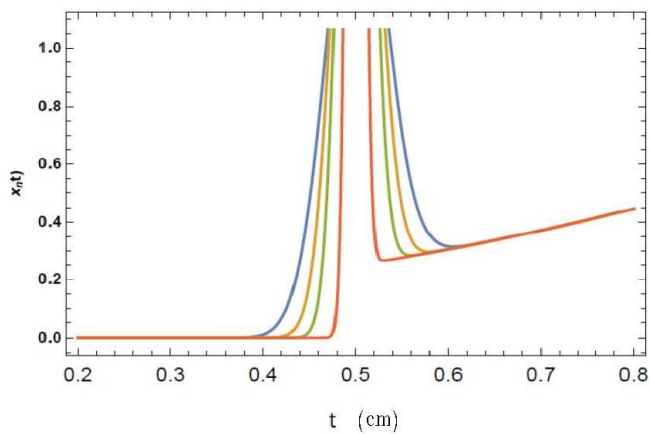}
\end{center}
\caption{Iterative solutions for the coordinate $x_{n}(t)$ for the
approximation of the acceleration $a^{(3)}_{n}(t)$. Notice how as the index
$n$ of the succession associated with the solution in the form of a
generalized function grows, the coordinates get closer and closer to the
expression (\ref{xf1}). The curves are reducing their width corresponding to
the growth of their index $n$ in the values $n = 500, 1000, 2000, 10000 $.
The time and distance are measured in natural units cm.}

\label{xnx}%
\end{figure}

As it can be seen in the figure \ref{vn}, for times different than the moment the force was applied $t=\frac{1}{2}$ and as the value of the
index $n$ grows, the elements of the succession of velocities $v_{n}%
(t),n=500,1000,2000,10000$ are getting closer and closer to the values of the
velocities of the solution (\ref{vf})%

\begin{equation}
v(t)=  t \,\, \overline{\theta}\left(  t-\frac{1}%
{2}\right)  . \label{vf1}%
\end{equation}

The same behavior can be seen in figure \ref{xnx} for the succession of
coordinates $x_{n}(t), n= 500, 1000, 2000, 10000$ which approximates the
coordinate values also given by the solution (\ref{xf})%

\begin{equation}
x(t)=\left(  \frac{t^{2}}{2}+\frac{1}{8}\right)  \overline{\theta
}\left(  t-\frac{1}{2}\right)  . \label{xf1}%
\end{equation}

\section{ Electron moving between the plates of a capacitor}

Next, we will proceed to compare our solution for the case of an
instantaneously applied constant force with the one obtained by A. Yaghjian in
the reference \cite{Yighjian}. In this discussion the author poses the problem
of a charged particle that accelerates between the parallel plates of a
capacitor that produces a constant electric field $E_{0}$. The particle is
released at $t=0$ from a capacitor plate and leaves the capacitor through a
small hole in the opposite plate at time $t=t_{2}$. The causal and controlled
solutions for acceleration and velocity proposed by Yaghjian for this problem
are as follows:%

\begin{equation}
{\nu}^{\prime}(t)=\frac{eE_{0}}{m}\left\{
\begin{tabular}
[c]{l}%
$0$ \ \ \ for $t<0$\\
$1$ \ \ \ for $0^{+}<t<t_{2}$\\
$0$ \ \ \ for $t_{2}^{+}<t$   %
\end{tabular}
\ ,\right.
\end{equation}

\begin{equation}
{\nu}(t)=\left\{
\begin{tabular}
[c]{l}%
$0$ \ \ \ \ \ \ \ \ \ \ \ \ \ \ \ \ for $t<0$\\
$\Delta\nu_{1}+\frac{eE_{0}}{m}t$ \ \ \ for $0^{+}<t<t_{2}$, \\
$\Delta\nu_{21}+\frac{eE_{0}}{m}$ \ \ \ for $t_{2}^{+}<t$
\end{tabular}
\right.
\end{equation}
where $\Delta\nu_{1}=\nu(0^{+})$, $\Delta\nu_{2}=\nu(t_{2})$ y $\Delta\nu
_{21}=\Delta\nu_{1} + \Delta\nu_{2}$. On the other hand, the causal and
controlled solutions previously presented by us for velocity and acceleration
for every instant of time except for the instant in which the force was applied becomes

\begin{equation}
a^{i}(t)=\left\{
\begin{tabular}
[c]{l}%
$0$ \ \ \ \ for $t<0$\\
$\frac{f^{i}}{m}$ \ \ \ for $t>0$%
\end{tabular}
\ ,\right.
\end{equation}

\begin{equation}
v^{i}(t)=\left\{
\begin{tabular}
[c]{l}%
$v^{i}(t_{0})$ \ \ \ \ \ \ \ \ \ \ \ \ \ \ \ \ \ \ \ for $t<0$\\
$v^{i}(t_{0})+\frac{f^{i}}{m}(t+\frac{k}{m})$ \ \ \ for $t>0$%
\end{tabular}
\ .\right.
\end{equation}

From here we see that if we take $f^{i}=eE_{0}$ and make $v^{i}(t_{0})=0$,
$\frac{f^{i}k}{m^{2}}=\Delta\nu_{1}$ then both solutions coincide for every
instant of time in which the constant force acts except for that instant in which the force was applied. This constitutes an important result if we take into account that
in the reference \cite{Yighjian} this conclusion is obtained after assuming
the finite spatial extension of the particle. Therefore, the coincidence
suggests the possible equivalence between the description presented in this
work and the one proposed by A. Yaghjian. Thus, we consider of great interest
to verify this possible equivalence in further works once the results obtained are extended to a relativistic context.

\section{ Solution for a constant magnetic field}

In this section we want to present the solution of the non-relativistic
Newtonian equation for the important case of a particle moving in a constant
magnetic field. Our objective is to verify that the solution of this problem
reproduces the movement of the particle in the same magnetic field but for the
non-relativistic ALD equation. The solution to this problem with the ALD
equations was found in the reference \cite{plass}, the resulting motion is
described by a spiral that tends to a point for sufficiently long times.

The solution that we will show below reproduces this behavior, then provides
another example of motion that simultaneously satisfies the Newtonian like equations
and the ALD ones \cite{jackson, zhang, PRD}.

Consider the force vector in the form $f^{i}(t)$ as the Lorentz force in a
constant magnetic field%

\begin{equation}
f^{i}(t)=q\epsilon^{ijk}v^{j}(t)B^{k},
\end{equation}
where the indices $i, j, k=1,2,3$, $\epsilon^{i, j, k}$ is the Levi-Civita
symbol, that stands for the vector product and $q$ is the charge of the particle. For
simplicity the magnetic field will be written in the form%

\begin{equation}
B^{k}=B\delta^{k3},
\end{equation}
where $\delta^{ki}$ is the Kronecker Delta with indices $k, i $.

After substituting the expression for the Lorentz force in the Newtonian
equations they take the form%

\begin{align}
a^{i}(t)  &  =\frac{1}{m}\sum_{l=0}^{\infty}\left(  \frac{\kappa}{m}\right)
^{l}\frac{d^{l}}{dt^{l}}f^{i}(t),\nonumber\\
\frac{d}{dt}v^{j}(t)  &  =\frac{qB}{m}\epsilon^{ij}\sum_{l=0}^{\infty} \left(
\frac{\kappa}{m}\right)  ^{l}\frac{d^{l}}{dt^{l}}v^{j}(t), \label{N}%
\end{align}
where the two-index tensor $\epsilon^{ij}$ is defined in terms of the
Levi-Civita symbol as%

\begin{equation}
\epsilon^{ij}=\epsilon^{ij3}.
\end{equation}

Let's separate the equation (\ref{N}) into two equations, one for the motion
in the plane orthogonal to the magnetic field and the other for the motion
parallel to it%

\begin{align}
\frac{d}{dt}v^{j}(t)  &  =\frac{qB}{m}\epsilon^{ij}\sum_{l=0}^{\infty} \left(
\frac{\kappa}{m}\right)  ^{l}\frac{d^{l}}{dt^{l}}v^{j}(t),\text{
}\ i=1,2,\label{Np}\\
\frac{d}{dt}v^{3}(t)  &  =0.
\end{align}

From here we observe that the movement parallel to the field is uniform%

\begin{align}
v^{3}(t)  &  =v_{0}^{3},\\
x^{3}(t)  &  =v_{0}^{3}t+x_{0}^{3}.
\end{align}

For the movement in the orthogonal plane will be useful to define the complex variable%

\begin{equation}
v(t)=v^{1}(t)+i \, v^{2}(t).
\end{equation}

But, since multiplying a complex number by the imaginary unit is equivalent to
rotating the corresponding 2D vector to $\frac{\pi}{2}$ from the $x_{1}$ axis towards the $x_{2}$ axis we can establish the following equivalence%

\begin{equation}
V^{\prime i}=\epsilon^{ij}V^{j}\text{ \ }\equiv\text{ \ \ \ }V^{\prime}=iV.
\end{equation}

Then, the equation for the motion in the plane orthogonal to the field
(\ref{Np}) can be written in a complex expression as follows%

\begin{align}
\frac{d}{dt}v(t)  &  =\frac{iqB}{m}\sum_{l=0}^{\infty}\tau^{l}\frac{d^{l}%
}{dt^{l}}v(t) ,\label{Npc}\\
\tau &  =\frac{\kappa}{m}.
\end{align}

Considering that the equation is invariant to a shift in time $t\rightarrow t+a$, we will look for exponential solutions with the form%

\begin{equation}
v(t)=\exp(iwt)v_{0}.
\end{equation}

Substituting this expression in (\ref{Npc}) we have%

\begin{align}
iwv_{0}  &  =\frac{iqB}{m}\sum_{l=0}^{\infty}(i\tau w)^{l}v_{0}\\
&  =\frac{iqB}{m}\frac{1}{1-i\tau w}v_{0}.
\end{align}

This formula can be written as%

\begin{equation}
\left(  w^{2}+i\frac{w}{\tau}-\frac{iqB}{\tau m}\right)  v_{0}=0.
\end{equation}

Therefore, this relation implies that the parameter $w$ satisfies%

\begin{equation}
\label{ser}w^{2}+i\frac{w}{\tau}-\frac{iqB}{\tau m}=0.
\end{equation}

The two solutions of this equation are%

\begin{equation}
w=-\frac{i}{2\tau}\pm\frac{i}{2\tau}\sqrt{1-\frac{4qB\tau}{m}i}.
\end{equation}

But, at the zero magnetic field limit the equation takes the form%

\begin{equation}
\frac{d}{dt}v(t)=iwv(t)=0,\nonumber
\end{equation}

implying that the physical solution of the equation (\ref{ser}) is%

\begin{equation}
\label{wer}w=-\frac{i}{2\tau}+\frac{i}{2\tau}\sqrt{1-\frac{4qB\tau}{m}i}.
\end{equation}

\begin{figure}[h]
\begin{center}
\includegraphics[width=0.4\textwidth]{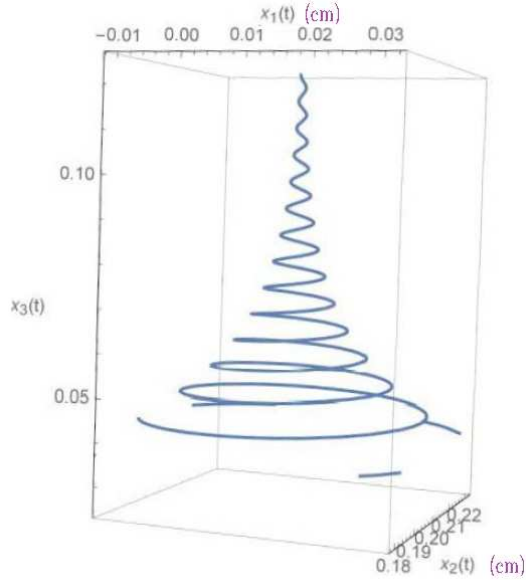}
\end{center}
\caption{The graph shows the resulting spiral motion for a charged particle in
a constant magnetic field. The movement in the direction parallel to the field
is uniform. The kinetic energy of the particle in the plane orthogonal to the
field is constantly radiated and the motion becomes uniform at the end. The
solution was found first in the reference \cite{plass} after solving the ALD
equation. The derivative here after solving the Newtonian equation was
obtained in order to verify the exact equivalence between both equations.
 The spatial and time axis units are in the natural system of units $cm$.}%
\label{xb}%
\end{figure}

The complex representation of the solution for the velocity is%

\begin{equation}
v(t)=\exp\left[  i\left(  -\frac{i}{2\tau}+\frac{i}{2\tau}\sqrt{1-\frac
{4qB\tau}{m}i}\right)  t\right]  v_{0}. \label{solComplex}%
\end{equation}

Therefore, for a real parameter $v_{0}$ the solutions for the velocity
components are%

\begin{align}
v^{1}(t)  &  =\operatorname{Re}\left\lbrace  \exp\left[  i\left(  -\frac{i}{2\tau}
+\frac{i}{2\tau}\sqrt{1-\frac{4qB\tau}{m}i}\right)  t\right]  \right\rbrace
v_{0},\\
v^{2}(t)  &  =\operatorname{Im}\left\lbrace  \exp\left[  i\left(  -\frac{i}{2\tau}
+\frac{i}{2\tau}\sqrt{1-\frac{4qB\tau}{m}i}\right)  t\right]  \right\rbrace
v_{0},\\
v^{3}(t)  &  =v_{0}^{3}.
\end{align}

The components of the coordinate are obtained after integrating the
expressions for the velocity%
\begin{align}
x^{1}(t)  &  =\operatorname{Re}\left\lbrace  \frac{\exp\left[  i\left(  -\frac
{i}{2\tau}+\frac{i}{2\tau}\sqrt{1-\frac{4qB\tau}{m}i}\right)  t\right]  -1}
{i\left(  -\frac{i}{2\tau}+\frac{i}{2\tau}\sqrt{1-\frac{4qB\tau}{m}i}\right)
}\right\rbrace  v_{0},\\
x^{2}(t)  &  =\operatorname{Im}\left\lbrace  \frac{\exp\left[  i\left(  -\frac
{i}{2\tau}+\frac{i}{2\tau}\sqrt{1-\frac{4qB\tau}{m}i}\right)  t\right]  -1}
{i\left(  -\frac{i}{2\tau}+\frac{i}{2\tau}\sqrt{1-\frac{4qB\tau}{m}i}\right)
}\right\rbrace  v_{0},\\
x^{3}(t)  &  =v_{0}^{3}t+x_{0}^{3}.
\end{align}

The evolution of the coordinates in time is shown in the figure \ref{xb}. In
this picture it can be observed how the particle radiates its kinetic energy
while it advances, which determines the spiral shape of the trajectory in the
plane perpendicular to the magnetic field.

\subsection{ Equivalence between Newtonian equations and ALD equations.}

Let us now show that the solution found satisfies the ALD equation as well. Then, the ALD equation associated to this case is:

\begin{equation}
a^{i}(t)-\frac{eB}{m}\epsilon^{ij}v^{j}(t)=\tau\frac{d}{dt}a^{i}(t),
\label{ALD11}%
\end{equation}

which in complex notation becomes%

\begin{equation}
a(t)-i\text{ }\frac{eB}{m}v(t)=\tau\frac{d}{dt}a(t).
\end{equation}

But, the derivative of the acceleration, after taking into account the
expression (\ref{wer}), becomes%

\begin{align}
\tau\frac{d}{dt}a(t)  &  =\tau\frac{d^{2}}{dt^{2}}v(t)\nonumber\\
&  =\tau(i\text{ }w)^{2}\exp(iwt)v_{0}\nonumber\\
&  =\tau(iw)^{2}v(t)\nonumber\\
&  =\left(  \frac{1}{2\tau}-i\frac{qB}{m}-\sqrt{1-\frac{4qB\tau}{m}i}\right)
v(t).
\end{align}

On the other hand, the left side of the ALD equation becomes%

\begin{align}
a(t)-i\frac{eB}{m}v(t)  &  =\frac{d}{dt}v(t)-i\frac{eB}{m}v(t)\nonumber\\
&  =\left(  iw-i\frac{eB}{m}\right)  v(t)\nonumber\\
&  =\left(  \frac{1}{2\tau}-i\frac{qB}{m}-\sqrt{1-\frac{4qB\tau}{m}i}\right)
v(t).
\end{align}

Therefore, the solution found also satisfies the ALD equation.

\section*{ Summary}

In the present work, we sought to formulate a causal and controlled solution for the non-relativistic Abraham-Lorentz-Dirac equation.
For this purpose, we focused on the search for a class of functions, which we called $P$, that would make the Landau-Lifshitz series convergent. Then, it turned out that the solutions formulated in the space spanned by such functions were equivalent to the solutions of the ALD equation.  The following results were obtained

\begin{enumerate}
\item A wide class of forces was identified, for which exact solutions to both, the ALD and the Newton-like equation, could be formulated. This set of forces is composed of polynomial functions of the time $t$ defined in an open interval of time. Also, these functions satisfy the Weierstrass Theorem, which states that any continuous or piecewise continuous function can be approximated with arbitrary precision in the aforementioned interval. Thus, almost any force can be represented by the elements of this class.

\item We argue the existence of a wider set of solutions common to the ALD and the Newton-like equations, in the terms of generalized functions which were defined by sequences of functions of $P$.  These solutions did not show any kind of non-causal behavior or uncontrolled growth.

\item An important solution in terms of generalized functions was found. It was the solution for the instantaneously applied force, which might be considered the most relevant result of this work. As it is known, it is in this case that the ALD equations show the most distinctive difficulties associated with this equation, namely, runaway and pre-accelerated behavior. It was
explicitly shown how the successions defining the generalized functions get
closer and closer to the solution for every instant different than the moment the force was applied when the index of the successions tends to infinity.

\item Another important conclusion of the discussion is that the solution exactly
reproduces the one obtained by A. Yaghjian \cite{Yighjian} for the problem of
a particle moving between the plates of a capacitor that produces a constant
electric field. This conclusion suggests the possible equivalence between the
description presented  and Yaghjian's
proposal in terms of the finite spatial extent of the particle.

\item Similarly, the non-relativistic Newtonian equation was applied to the
important case of a charged particle moving in a constant magnetic field, the
solution obtained reproduces the particle's motion determined by G. N. Plass
in \cite{plass}. In that reference, the solution was derived after solving the
ALD equation for this case. The motion derived here was shown to
simultaneously satisfy the Newton-like as well as the ALD equations.
\end{enumerate}

The conclusions drawn from this work, as well as the information acquired and
questions that arose during the writing process, motivated new objectives to
direct our efforts. A few points of interest for future studies are:

\begin{itemize}
\item To extend the study to the relativistic case, thus formulating a more
general discussion that allows verifying that the solution obtained here
constitutes a particular limit case of a more general one

\item Afterwards, to study how the relativistic solutions allows to explain
why, as it was seen in this work, a discontinuous jump appear in the initial conditions for the coordinates of the particle, for the instantaneously applied constant force. The idea is that the relativistic solutions can show a continuous time variation of the coordinates that translates into discontinuous variation of the coordinates once the speed of light is taken in the infinite limit.

\item  The most ambitious result that is expected  to be obtained is to prove that a relativistic version of the Newton-like equation would be equivalent to the corresponding relativistic version of the ALD equations, yet avoiding the aforementioned drawbacks.

\item Investigate, the possible equivalence between the analysis presented
here and the one presented by A. Yaghjian in the reference \cite{Yighjian}.
\end{itemize}

\end{document}